\documentclass[envcountsame]{llncs}

\usepackage{amssymb}
\title{{\bf On 
the Accepting Power of  \\ $2$-Tape B\"uchi Automata} }

\author{Olivier Finkel\inst{}}
\institute{Equipe de Logique Math\'ematique \\
 U.F.R. de Math\'ematiques, Universit\'e Paris 7 \\ 2 Place Jussieu 75251 Paris
 cedex 05, France \\ \email{finkel@logique.jussieu.fr}.}

\date{}

\begin{document}

\spnewtheorem{Rem}[theorem]{Remark}{\bfseries}{\itshape}
\spnewtheorem{Exa}[theorem]{Example}{\bfseries}{\itshape}

\spnewtheorem{Pro}[theorem]{Proposition}{\bfseries}{\itshape}
\spnewtheorem{Lem}[theorem]{Lemma}{\bfseries}{\itshape}
\spnewtheorem{Cor}[theorem]{Corollary}{\bfseries}{\itshape}
\spnewtheorem{Deff}[theorem]{Definition}{\bfseries}{\itshape}

\newcommand{\fa}{\forall}
\newcommand{\Ga}{\Gamma}
\newcommand{\Gas}{\Gamma^\star}
\newcommand{\Gao}{\Gamma^\omega}

\newcommand{\Si}{\Sigma}
\newcommand{\Sis}{\Sigma^\star}
\newcommand{\Sio}{\Sigma^\omega}
\newcommand{\ra}{\rightarrow}
\newcommand{\hs}{\hspace{12mm}

\noi}
\newcommand{\lra}{\leftrightarrow}
\newcommand{\la}{language}
\newcommand{\ite}{\item}
\newcommand{\Lp}{L(\varphi)}
\newcommand{\abs}{\{a, b\}^\star}
\newcommand{\abcs}{\{a, b, c \}^\star}
\newcommand{\ol}{ $\omega$-language}
\newcommand{\orl}{ $\omega$-regular language}
\newcommand{\om}{\omega}
\newcommand{\nl}{\newline}
\newcommand{\noi}{\noindent}
\newcommand{\tla}{\twoheadleftarrow}
\newcommand{\de}{deterministic }
\newcommand{\proo}{\noi {\bf Proof.} }
\newcommand {\ep}{\hfill $\square$}

\maketitle

\begin{abstract}
\noi 
We show that, from a topological point of view,  $2$-tape B\"uchi automata  have the same accepting power than 
Turing machines equipped with a  B\"uchi acceptance condition. 
In particular,   for every non null recursive ordinal $\alpha$,  
there exist some  
${\bf \Si}^0_\alpha$-complete and some ${\bf \Pi}^0_\alpha$-complete
infinitary rational relations accepted by $2$-tape B\"uchi automata.  This  surprising result gives answers to 
questions of Simonnet \cite{sim} and of Lescow and Thomas \cite{tho,lt}. 
\end{abstract}

\noi {\small {\bf Keywords:} $2$-tape B\"uchi automata; infinitary rational relations;  Cantor topology; topological complexity; 
Borel hierarchy; complete sets.}

\section{Introduction}

In the sixties, automata accepting  infinite words were  firstly 
considered by B\"uchi in order to study 
decidability of the monadic second order theory S1S
of one successor over the integers \cite{bu62}. 
Then the so called \orl s have been 
intensively studied and have found many applications for 
specification and verification of non terminating systems, see \cite{tho,sta,pp} for many results and references. 
 On the other hand, rational relations on finite words were  also studied in the sixties,  and played 
a fundamental role in the study of families of context free languages \cite{ber}. 
Investigations on their extension to rational  relations on infinite words were carried out 
or mentioned in the books \cite{bt,ls}. Gire  and Nivat 
studied infinitary rational relations in  \cite{gire1,gn}. 
These relations  are sets of pairs of infinite words  which are accepted 
by $2$-tape finite B\"uchi automata with asynchronous 
reading heads. 
The class of infinitary rational relations, which  extends both the 
class of finitary rational relations  and  the class of \orl s,  and 
the rational functions they may define, have  been much studied, 
see for example  \cite{cg,bcps,sim,sta,pri}. 
\nl Notice that a  rational relation $R\subseteq \Si_1^\om \times \Si_2^\om$ may be seen as 
an \ol~ over the alphabet $\Si_1 \times \Si_2$.  
\nl  A way to study the complexity of languages of infinite words  
accepted by finite machines is to study their topological complexity and firstly
to locate them with regard to 
the Borel and the projective hierarchies. 
This work is analysed 
for example in \cite{stac,tho,eh,lt,sta}.  
It is well known that every \ol~ accepted by a Turing machine with a 
B\"uchi or Muller acceptance condition is an analytic set and 
 that \orl s are boolean combinations of ${\bf \Pi}_2^0$-sets 
hence ${\bf \Delta}_3^0$-sets,  \cite{sta,pp}.  
\nl  The question of the topological complexity of  relations on infinite words also 
naturally arises and is asked by Simonnet in \cite{sim}. It is also posed in a more 
general form by Lescow and Thomas in \cite{lt} 
(for infinite labelled partial orders) and in \cite{tho89} 
where Thomas suggested to study reducibility notions and associated completeness results.  
\nl Every infinitary rational relation is an analytic set. 
We showed in \cite{relrat} that there exist some infinitary rational relations 
which are analytic but non  Borel, and in \cite{relratbor} that there are some ${\bf \Si}^0_3$-complete 
and some ${\bf \Pi}^0_3$-complete infinitary rational relations, using a  coding of 
$\om^2$-words by pairs of infinite words. 
Using a different coding we proved in \cite{relratbor2} that there exist such 
infinitary rational relations which have  a very simple structure and 
can be easily described by their sections.  Using this very simple structure, 
we constructed also some infinitary rational relations,  accepted 
by 3-tape B\"uchi automata, 
which are ${\bf \Si}^0_4$-complete. 
\nl On the other hand we recently proved in \cite{cie,mscs} that the Borel hierarchy of  $\om$-languages accepted by 
B\"uchi  real time $1$-counter automata is equal to the Borel hierarchy of  $\om$-languages accepted by B\"uchi Turing machines. 
In particular, for each  non null recursive ordinal $\alpha$,  
there exist some  
${\bf \Si}^0_\alpha$-complete and some ${\bf \Pi}^0_\alpha$-complete $\om$-languages accepted by 
B\"uchi real time $1$-counter automata. 
\nl Using a simulation of real time $1$-counter automata we prove in this paper a similar result:   
 the Borel hierarchy of the class of infinitary rational relations is equal to the Borel hierarchy of
$\om$-languages accepted by B\"uchi real time $1$-counter automata which is also equal to the 
Borel hierarchy of $\om$-languages accepted by B\"uchi Turing machines. In particular, 
 for each  non null recursive ordinal $\alpha$,  there exist some  
${\bf \Si}^0_\alpha$-complete and some ${\bf \Pi}^0_\alpha$-complete infinitary rational relations. This 
gives  answers to questions of Simonnet \cite{sim} and of Lescow and Thomas \cite{tho,lt}. 
\nl  The paper is organized as follows. In section 2 we recall  
the notion of  2-tape  automata and of real time $1$-counter automata with B\"uchi acceptance condition. 
In section 3 we recall definitions of Borel and analytic
sets, and we prove our main result  in section 4.

\section{$2$-tape automata and $1$-counter automata}

\noi We assume the reader to be familiar with the theory of formal ($\om$)-languages  
\cite{tho,sta}.
We shall use usual notations of formal language theory. 
\nl  When $\Si$ is a finite alphabet, a {\it non-empty finite word} over $\Si$ is any 
sequence $x=a_1\ldots a_k$, where $a_i\in\Sigma$ 
for $i=1,\ldots ,k$ , and  $k$ is an integer $\geq 1$. The {\it length}
 of $x$ is $k$, denoted by $|x|$.
 The {\it empty word} has no letter and is denoted by $\lambda$; its length is $0$. 
 For $x=a_1\ldots a_k$, we write $x(i)=a_i$  
and $x[i]=x(1)\ldots x(i)$ for $i\leq k$ and $x[0]=\lambda$.
 $\Sis$  is the {\it set of finite words} (including the empty word) over $\Sigma$.
 \nl  The {\it first infinite ordinal} is $\om$.
 An $\om$-{\it word} over $\Si$ is an $\om$ -sequence $a_1 \ldots a_n \ldots$, where for all 
integers $ i\geq 1$, ~
$a_i \in\Sigma$.  When $\sigma$ is an $\om$-word over $\Si$, we write
 $\sigma =\sigma(1)\sigma(2)\ldots \sigma(n) \ldots $,  where for all $i$,~ $\sigma(i)\in \Si$,
and $\sigma[n]=\sigma(1)\sigma(2)\ldots \sigma(n)$  for all $n\geq 1$ and $\sigma[0]=\lambda$.
\nl   The {\it prefix relation} is denoted $\sqsubseteq$: a finite word $u$ is a {\it prefix} 
of a finite word $v$ (respectively,  an infinite word $v$), denoted $u\sqsubseteq v$,  
 if and only if there exists a finite word $w$ 
(respectively,  an infinite word $w$), such that $v=u.w$.
 The {\it set of } $\om$-{\it words} over  the alphabet $\Si$ is denoted by $\Si^\om$.
An  $\om$-{\it language} over an alphabet $\Sigma$ is a subset of  $\Si^\om$.  The complement (in $\Sio$) of an 
$\om$-language $V \subseteq \Sio$ is $\Sio - V$, denoted $V^-$.
 
\hs 
Infinitary rational relations 
are subsets of  $\Sio \times \Gao$,  where 
$\Si$ and  $\Ga$ are finite alphabets, which are accepted by 
$2$-tape B\"uchi automata  (2-BA). 

\begin{Deff}
A  2-tape B\"uchi automaton 
 is a sextuple $\mathcal{T}=(K, \Si, \Ga, \Delta, q_0, F)$, where 
$K$ is a finite set of states, $\Si$ and $\Ga$ are finite  alphabets, 
$\Delta$ is a finite subset of $K \times \Sis \times \Gas \times K$ called 
the set of transitions, $q_0$ is the initial state,  and $F \subseteq K$ is the set of 
accepting states. 
\nl A computation $\mathcal{C}$ of the  
2-tape B\"uchi automaton $\mathcal{T}$ is an infinite sequence of transitions 
$$(q_0, u_1, v_1, q_1), (q_1, u_2, v_2, q_2), \ldots (q_{i-1}, u_{i}, v_{i}, q_{i}), 
(q_i, u_{i+1}, v_{i+1}, q_{i+1}), \ldots $$
\noi The computation is said to be successful iff there exists a final state $q_f \in F$ 
and infinitely many integers $i\geq 0$ such that $q_i=q_f$. 
\nl The input word of the computation is $u=u_1.u_2.u_3 \ldots$
\nl The output word of the computation is $v=v_1.v_2.v_3 \ldots$
\nl Then the input and the output words may be finite or infinite. 
\nl The infinitary rational relation $R(\mathcal{T})\subseteq \Sio \times \Ga^\om$ 
accepted by the 2-tape B\"uchi automaton $\mathcal{T}$ 
is the set of couples $(u, v) \in \Sio \times \Ga^\om$ such that $u$ and $v$ are the input 
and the output words of some successful computation $\mathcal{C}$ of $\mathcal{T}$. 
\nl The set of infinitary rational relations will be denoted $RAT_\om$. 
\end{Deff} 

\begin{Deff} 
A   ({\it real time})  $1$-counter machine  is a 4-tuple 
$\mathcal{M}$=$(K,\Si, \Delta, q_0)$,  where $K$ 
is a finite set of states, $\Sigma$ is a finite input alphabet, 
 $q_0\in K$ is the initial state, 
and the transition relation $\Delta$ is a subset of  
$K \times  \Si   \times \{0, 1\} \times K \times \{0, 1, -1\}$. 
\nl  
If  the machine $\mathcal{M}$ is in state $q$ and 
$c \in \mathbf{N}$ is the content of the counter  then 
the  configuration (or global state)
 of $\mathcal{M}$ is  $(q, c)$.

\hs For $a\in \Si $, 
$q, q' \in K$ and $c \in \mathbf{N}$, if 
$(q, a, i, q', j) \in \Delta$ where $i=0$ if $c=0$ 
and $i=1$ if $c\neq 0$  then we write:
$$a: (q, c)\mapsto_{\mathcal{M}} (q', c+j)$$

\noi $\mapsto_{\mathcal{M}}^\star$ is the transitive and reflexive closure of
 $\mapsto_{\mathcal{M}}$.
\nl Thus we see that the transition relation must satisfy:
 \nl if $(q, a,  i, q', j)   \in   \Delta$ and  $i=0$ then $j=0$ or $j=1$ (but $j$ may not be equal to $-1$).

\hs
Let $\sigma =a_1a_2 \ldots a_n $ be a finite word over $\Si$. 
A sequence of configurations $r=(q_i, c_{i})_{1\leq i \leq n+1}$  is called 
a  run of $\mathcal{M}$ on $\sigma$, starting in configuration 
$(p, c)$, iff:
\begin{enumerate}
\ite[(1)]  $(q_1, c_{1})=(p, c)$
\ite[(2)] 
 for each $i \in [1, n]$, 
 $a_i: (q_i, c_{i})\mapsto_{\mathcal{M}}
(q_{i+1},  c_{i+1})$ 
\end{enumerate}

\noi 
Let $\sigma =a_1a_2 \ldots a_n \ldots $ be an $\om$-word over $\Si$. 
An $\om$-sequence of configurations $r=(q_i, c_{i})_{i \geq 1}$ is called 
a run of $\mathcal{M}$ on $\sigma$, starting in configuration 
$(p, c)$, iff:
\begin{enumerate}
\ite[(1)]  $(q_1, c_{1})=(p, c)$

\ite[(2)]   for each $i\geq 1$, 
 $a_i: (q_i, c_{i})\mapsto_{\mathcal{M}}  
(q_{i+1},  c_{i+1})$  
\end{enumerate}
\noi
For every such run, $\mathrm{In}(r)$ is the set of all states entered infinitely
 often during run $r$.
\nl
A  run $r$ of $M$ on $\sigma$, starting in configuration $(q_0, 0)$,
 will be simply called ``a run of $M$ on $\sigma$".
\end{Deff}

\begin{Deff} A ({\it real time}) B\"uchi $1$-counter automaton  is a 5-tuple 
\begin{center} 
$\mathcal{M}$=$(K,\Si, \Delta, q_0, F), $
\end{center} 
where $ \mathcal{M}'$=$(K,\Si, \Delta, q_0)$
is a ({\it real time}) $1$-counter machine and $F \subseteq K$ 
is the set of accepting  states.
The \ol~ accepted by $\mathcal{M}$ is 
\begin{center}
$L(\mathcal{M})$= $\{  \sigma\in\Si^\om \mid \mbox{  there exists a  run r
 of } \mathcal{M} \mbox{ on } \sigma \mbox{  such that } \mathrm{In}(r)
 \cap F \neq \emptyset \}$
\end{center}
\end{Deff}

\noi The class of  (real time)  B\"uchi $1$-counter automata   will be 
denoted {\bf r}-${\bf BC}(1)$.
\nl The class of \ol s accepted by real time B\"uchi $1$-counter automata  will be 
denoted {\bf r}-${\bf BCL}(1)_\om$.

\section{Borel hierarchy}

\noi We assume the reader to be familiar with basic notions of topology which
may be found in \cite{mos,lt,kec,sta,pp}.
There is a natural metric on the set $\Sio$ of  infinite words 
over a finite alphabet 
$\Si$ which is called the {\it prefix metric} and defined as follows. For $u, v \in \Sio$ and 
$u\neq v$ let $\delta(u, v)=2^{-l_{\mathrm{pref}(u,v)}}$ where $l_{\mathrm{pref}(u,v)}$ 
 is the first integer $n$
such that the $(n+1)^{st}$ letter of $u$ is different from the $(n+1)^{st}$ letter of $v$. 
This metric induces on $\Sio$ the usual  Cantor topology for which {\it open subsets} of 
$\Sio$ are in the form $W.\Si^\om$, where $W\subseteq \Sis$.
A set $L\subseteq \Si^\om$ is a {\it closed set} iff its complement $\Si^\om - L$ 
is an open set.
Define now the {\it Borel Hierarchy} of subsets of $\Si^\om$:

\begin{Deff}
For a non-null countable ordinal $\alpha$, the classes ${\bf \Si}^0_\alpha$
 and ${\bf \Pi}^0_\alpha$ of the Borel Hierarchy on the topological space $\Si^\om$ 
are defined as follows:
\nl ${\bf \Si}^0_1$ is the class of open subsets of $\Si^\om$, 
 ${\bf \Pi}^0_1$ is the class of closed subsets of $\Si^\om$, 
\nl and for any countable ordinal $\alpha \geq 2$: 
\nl ${\bf \Si}^0_\alpha$ is the class of countable unions of subsets of $\Si^\om$ in 
$\bigcup_{\gamma <\alpha}{\bf \Pi}^0_\gamma$.
 \nl ${\bf \Pi}^0_\alpha$ is the class of countable intersections of subsets of $\Si^\om$ in 
$\bigcup_{\gamma <\alpha}{\bf \Si}^0_\gamma$.
\end{Deff}

\noi For 
a countable ordinal $\alpha$,  a subset of $\Si^\om$ is a Borel set of {\it rank} $\alpha$ iff 
it is in ${\bf \Si}^0_{\alpha}\cup {\bf \Pi}^0_{\alpha}$ but not in 
$\bigcup_{\gamma <\alpha}({\bf \Si}^0_\gamma \cup {\bf \Pi}^0_\gamma)$.

\hs There are also some subsets of $\Si^\om$ which are not Borel.  In particular 
the class of Borel subsets of $\Si^\om$ is strictly included into 
the class  ${\bf \Si}^1_1$ of {\it analytic sets} which are 
obtained by projection of Borel sets, 
see for example \cite{sta,lt,pp,kec}
 for more details. 
The (lightface) class $\Si^1_1$ of {\it effective analytic sets} 
is the class of sets which are obtained by projection of arithmetical sets. It is 
well known that a set $L \subseteq \Sio$, where $\Si$ is a finite alphabet, 
 is in the class $\Si^1_1$  iff it is accepted by a Turing machine with a B\"uchi or Muller 
acceptance condition \cite{sta}. 
\nl  We now define completeness with regard to reduction by continuous functions. 
For a countable ordinal  $\alpha\geq 1$, a set $F\subseteq \Si^\om$ is said to be 
a ${\bf \Si}^0_\alpha$  
(respectively,  ${\bf \Pi}^0_\alpha$, ${\bf \Si}^1_1$)-{\it complete set} 
iff for any set $E\subseteq Y^\om$  (with $Y$ a finite alphabet): 
 $E\in {\bf \Si}^0_\alpha$ (respectively,  $E\in {\bf \Pi}^0_\alpha$,  $E\in {\bf \Si}^1_1$) 
iff there exists a continuous function $f: Y^\om \ra \Si^\om$ such that $E = f^{-1}(F)$. 
 ${\bf \Si}^0_n$
 (respectively ${\bf \Pi}^0_n$)-complete sets, with $n$ an integer $\geq 1$, 
 are thoroughly characterized in \cite{stac}.  

\section{Topology and infinitary rational relations}

\noi    The first 
non-recursive ordinal, usually  called the Church-Kleene ordinal, will be denoted below by $\om_1^{\mathrm{CK}}$. 

\hs We have proved in \cite{cie,mscs} the following result. 

\begin{theorem}\label{thebor}    

\noi 
  For every non null countable ordinal $\alpha < \om_1^{\mathrm{CK}}$, 
there exist some  
${\bf \Si}^0_\alpha$-complete and some ${\bf \Pi}^0_\alpha$-complete
$\om$-languages in the class {\bf r}-${\bf BCL}(1)_\om$.  

\end{theorem}

\noi We are going to prove a similar  result for the class $RAT_\om$, 
 using a simulation of  $1$-counter automata. 
\nl We now first  define a coding of  an $\om$-word over a finite alphabet $\Si$ 
by an  $\om$-word over the  alphabet $\Ga = \Si \cup \{A\}$, where  $A$ is an additionnal letter 
not in $\Si$. 

\hs For $x\in \Sio$  the $\om$-word $h(x)$ is defined by : 
$$h(x) = A.0.x(1).A.0^2.x(2).A.0^3.x(3).A.0^4.x(4).A \ldots A.0^n.x(n).A.0^{n+1}.x(n+1).A \ldots$$
\noi Then it is easy to see that the mapping $h$ from $\Sio$ into $(\Si \cup \{A\})^\om$ is continuous and injective. 

\begin{Lem}\label{lem}
Let $\Si$ be a finite alphabet and $\alpha\geq 2$ be a countable ordinal. 
If $L \subseteq \Si^{\om}$ is  
${\bf \Pi}^0_\alpha$-complete (respectively,    ${\bf \Si}^0_\alpha$-complete)            
  then 
$$h(L) \cup h(\Si^{\om})^-$$
\noi  is a ${\bf \Pi}^0_\alpha$-complete  (respectively,    ${\bf \Si}^0_\alpha$-complete)    
subset of $(\Si \cup\{A\})^\om$.
\end{Lem}

\proo 
  Let $L$ be a ${\bf \Pi}^0_\alpha$-complete (respectively,    ${\bf \Si}^0_\alpha$-complete)   
subset of $\Sio$,  for some countable ordinal $\alpha\geq 2$ . 

\hs The topological space $\Si^{\om}$  is compact 
thus its image by the continuous function 
$h$ is also a compact subset of the topological space 
$(\Si \cup\{A\})^\om$. 
The set  $h(\Si^{\om})$ is compact hence  it is a closed subset of 
$(\Si \cup\{A\})^\om$ and  its complement 
$$(h(\Si^{\om}))^- =  (\Si \cup\{A\})^\om - h(\Si^{\om})$$
\noi  is an open (i.e. a ${\bf \Si}^0_1$) subset of $(\Si \cup\{A\})^\om$.

\hs  On the other side the function $h$ is also injective 
thus it is a bijection from $\Si^{\om}$  onto 
$h(\Si^{\om})$. But a continuous bijection between two compact sets is an homeomorphism
therefore $h$ induces an homeomorphism between  $\Si^{\om}$  and  $h(\Si^{\om})$. 
By hypothesis $L$ is a  ${\bf \Pi}^0_\alpha$ (respectively,    ${\bf \Si}^0_\alpha$)-subset    of $\Si^{\om}$  thus 
$h(L)$ is a 
${\bf \Pi}^0_\alpha$ (respectively,    ${\bf \Si}^0_\alpha$)-subset of 
 $h(\Si^{\om})$ (where Borel sets of the topological 
space $h(\Si^{\om})$ are defined from open sets as in the case of the topological 
space $\Sio$). 

\hs  The topological space $h(\Si^{\om})$ is a 
topological subspace of $ (\Si \cup\{A\})^\om$ and its 
topology  is induced by the topology on $(\Si \cup\{A\})^\om$: open sets 
of $h(\Si^{\om})$ are traces on $h(\Si^{\om})$ of open sets of 
$(\Si \cup\{A\})^\om$ and the same result holds for closed sets. Then 
one can easily show by induction that for every ordinal   $\beta \geq 1$,  
${\bf \Pi}^0_\beta $-subsets 
(resp.  ${\bf \Si}^0_\beta $-subsets) of  $h(\Si^{\om})$ are traces on $h(\Si^{\om})$ of 
${\bf \Pi}^0_\beta$-subsets 
(resp.  ${\bf \Si}^0_\beta$-subsets) of  $(\Si \cup\{A\})^\om$, i.e. are 
intersections with $h(\Si^{\om})$ of 
${\bf \Pi}^0_\beta $-subsets 
(resp.  ${\bf \Si}^0_\beta$-subsets) of  $(\Si \cup\{A\})^\om$. 

\hs  But  $h(L)$ is a ${\bf \Pi}^0_\alpha$ (respectively,    ${\bf \Si}^0_\alpha$)-subset 
of  $h(\Si^{\om})$ hence there exists 
a  ${\bf \Pi}^0_\alpha$ (respectively,    ${\bf \Si}^0_\alpha$)-subset $T$ of  $(\Si \cup\{A\})^\om$ such that 
$h(L)=T \cap h(\Si^{\om})$. But $h(\Si^{\om})$ is a closed 
i.e. ${\bf \Pi}^0_1$-subset (hence also a ${\bf \Pi}^0_\alpha$ (respectively,    ${\bf \Si}^0_\alpha$)-subset) 
of $(\Si \cup\{A\})^\om$  and the class of ${\bf \Pi}^0_\alpha$ (respectively,    ${\bf \Si}^0_\alpha$)-subsets of 
$(\Si \cup\{A\})^\om$  is closed under finite intersection thus 
$h(L)$ is a ${\bf \Pi}^0_\alpha$ (respectively,    ${\bf \Si}^0_\alpha$)-subset of $(\Si \cup\{A\})^\om$.  
 
\hs  Now  $h(L) \cup  (h(\Si^{\om}))^-$ is the union of a ${\bf \Pi}^0_\alpha$ (respectively,  
  ${\bf \Si}^0_\alpha$)-subset
and of a ${\bf \Si}^0_1$-subset of $(\Si \cup\{A\})^\om$ 
therefore it is a ${\bf \Pi}^0_\alpha$ (respectively,    ${\bf \Si}^0_\alpha$)-subset of $(\Si \cup\{A\})^\om$ 
because the class of 
${\bf \Pi}^0_\alpha$ (respectively,    ${\bf \Si}^0_\alpha$)-subsets of $(\Si \cup\{A\})^\om$ 
 is closed under finite union.  

\hs  In order to prove that $h(L) \cup  (h(\Si^{\om}))^-$ 
 is ${\bf \Pi}^0_\alpha$ (respectively,    ${\bf \Si}^0_\alpha$)-{\bf-complete} it suffices to remark 
that
$$L=h^{-1}[ h(L) \cup  (h(\Si^{\om}))^- ]$$
\noi This implies that $h(L) \cup  (h(\Si^{\om}))^- $ is ${\bf \Pi}^0_\alpha$ 
(respectively,    ${\bf \Si}^0_\alpha$)-complete 
because  $L$ is assumed to be ${\bf \Pi}^0_\alpha$ 
(respectively,    ${\bf \Si}^0_\alpha$)-complete.    \ep

\hs Let now $\Si$ be a finite alphabet such that $0\in \Si$ and let $\alpha$ be the $\om$-word over the alphabet 
$\Si \cup\{A\}$ which is defined by:

$$\alpha = A.0.A.0^2.A.0^3.A.0^4.A.0^5.A \ldots A.0^n.A.0^{n+1}.A \ldots$$

\noi We can now state the following Lemma.

\begin{Lem}\label{R1}
Let $\Si$  be a finite alphabet such that $0\in \Si$, 
$\alpha$ be the $\om$-word over  $\Si \cup\{A\}$ defined as above, and 
 $L \subseteq \Sio$ be in  {\bf r}-${\bf BCL}(1)_\om$.
Then there exists  an infinitary rational relation 
$R_1 \subseteq (\Si\cup\{A\})^\om \times (\Si \cup\{A\})^\om$ such that:
$$\fa x\in \Si^{\om}~~~ (x\in L) \mbox{  iff } ( (h(x), \alpha) \in R_1 )$$ 
\end{Lem}

\proo Let $\Si$  be a finite alphabet such that $0\in \Si$, $\alpha$ be the $\om$-word 
over  $\Si \cup\{A\}$ defined as above, and  $L = L(\mathcal{A}) \subseteq \Sio$, where 
$\mathcal{A}$=$(K,\Si, \Delta, q_0, F)$ is a $1$-counter B\"uchi automaton. 

\hs We define now the relation $R_1$.
A pair 
$y=(y_1, y_2)$ of $\om$-words over the alphabet $\Si\cup\{A\}$ is in $R_1$ 
if and only if it is in the form

\hs $y_1 = A.u_1.v_1.x(1).A.u_2.v_2.x(2).A.u_3.v_3.x(3).A  
\ldots A.u_{n}.v_{n}.x(n).A. \ldots$
\nl $y_2 = A.w_1.z_1.A.w_2.z_2.A.w_3.z_3.A \ldots 
 A.w_{n}.z_{n}.A \ldots$

\hs where $|v_1|=0$ and 
for all integers $i\geq 1$, 
$$ u_i ,v_i, w_i, z_i \in 0^\star \mbox{ and } x(i) \in \Si  \mbox{  and   }  $$ 
$$~~~~~ |u_{i+1}|=|z_i|+1 $$ 

\noi and there is a sequence $(q_i)_{i\geq 0}$  of states of $K$ 
 such that  for all integers 
$i\geq 1$:  

$$  x(i) : ( q_{i-1}, |v_i| ) \mapsto_{\mathcal{A}} 
(q_i, |w_i| )$$

\noi Moreover some state $q_f \in F$ occurs infinitely often in the sequence $(q_i)_{i\geq 0}$. 
\nl  Notice that the state $q_0$ of the sequence $(q_i)_{i\geq 0}$  is also the initial state 
of $\mathcal{A}$. 

\hs Let now $x\in \Si^{\om}$ such that  $  (h(x), \alpha) \in R_1$. We are going to prove that $x\in L$. 

\hs  By hypothesis  $ (h(x), \alpha) \in R_1$ thus there are finite words  $u_i ,v_i, w_i, z_i \in 0^\star$ such that 
 $|v_1|=0$ and for all integers $i\geq 1$, $|u_{i+1}|=|z_i|+1 $, and  

\hs $h(x) = A.u_1.v_1.x(1).A.u_2.v_2.x(2).A.u_3.v_3.x(3).A  
\ldots A.u_{n}.v_{n}.x(n).A. \ldots$

\hs $\alpha  = A.w_1.z_1.A.w_2.z_2.A.w_3.z_3.A \ldots 
 A.w_{n}.z_{n}.A \ldots$

\hs Moreover  there is a sequence $(q_i)_{i\geq 0}$  of states of $K$ 
 such that  for all integers 
$i\geq 1$:  

$$  x(i) : ( q_{i-1}, |v_i| ) \mapsto_{\mathcal{A}} 
(q_i, |w_i| )$$

\noi and  some state $q_f \in F$ occurs infinitely often in the sequence $(q_i)_{i\geq 0}$. 

\hs 
on the other side we have: 
\nl  $h(x) = A.0.x(1).A.0^2.x(2).A.0^3.x(3).A \ldots A.0^n.x(n).A.0^{n+1}.x(n+1).A \ldots$
\nl $\alpha =A.0.A.0^2.A.0^3.A.0^4.A  \ldots A.0^n.A \ldots$

\hs So we have $|u_1.v_1|=1$ and $|v_1|=0$ and $x(1): ( q_{0}, |v_1| ) \mapsto_{\mathcal{A}} 
(q_1, |w_1| )$. But $|w_1.z_1|=1$,  $|u_2.v_2|=2$, and $|u_2|=|z_1|+1$ thus $|v_2|=|w_1|$. 

\hs We are going to prove  in a similar way  that for all integers $i\geq 1$ it holds that $|v_{i+1}|=|w_i|$. 
\nl We know that $|w_i.z_i|=i$, $|u_{i+1}.v_{i+1}|=i+1$, and $|u_{i+1}|=|z_i|+1$ thus $|w_i|=|v_{i+1}|$. 

\hs Then for all $i\geq 1$,  $x(i) : ( q_{i-1}, |v_i| ) \mapsto_{\mathcal{A}} (q_i, |v_{i+1}| )$. 
\nl So if we set $c_i=|v_i|$, $(q_{i-1}, c_{i})_{i\geq 1}$ is an accepting run of $\mathcal{A}$ on $x$  and this implies that  
$x\in L$. 
\nl Conversely it is easy to prove that if $x\in L$ then $(h(x), \alpha)$ may be written in the form of $(y_1, y_2)\in R_1$. 

\hs It remains  to prove that the above defined relation $R_1$ is an infinitary rational 
relation.  It is easy to find a $2$-tape  B\"uchi automaton $\mathcal{T}$ accepting  
the infinitary rational relation $R_1$.

\begin{Lem}\label{complement} The set 
$$R_2 = (\Si\cup\{A\})^\om\times (\Si \cup\{A\})^\om - ( h(\Si^{\om}) \times \{\alpha\} )$$
\noi is an infinitary rational relation.  
\end{Lem}

\proo By definition of the mapping $h$, we know that a pair of $\om$-words 
over   the alphabet  $(\Si\cup\{A\})$ is in $h(\Si^{\om}) \times \{\alpha\}$ iff 
 it is  in the form $(\sigma_1, \sigma_2)$, 
where
\hs  $\sigma_1 = A.0.x(1).A.0^2.x(2).A.0^3.x(3).A \ldots 
.A.0^n.x(n).A.0^{n+1}.x(n+1).A \ldots$
\nl  $\sigma_2 = \alpha = A.0.A.0^2.A.0^3.A \ldots A.0^n.A.0^{n+1}.A \ldots$

\hs where for all integers $i\geq 1$,  $x(i)\in \Si$. 

\hs So it is easy to see that 
$(\Si\cup\{A\})^\om\times (\Si \cup\{A\})^\om - ( h(\Si^{\om}) \times \{\alpha\} )$
is the union of the sets $\mathcal{C}_j$ where:

\begin{itemize} 

\ite $\mathcal{C}_1$ is formed by pairs  $(\sigma_1, \sigma_2)$ where 
\nl $\sigma_1$ has not any initial segment in $A.\Si^2.A.\Si^3.A$, 
or $\sigma_2$ has not any initial segment in $A.\Si.A.\Si^2.A$.

\ite $\mathcal{C}_2$ is formed by pairs  $(\sigma_1, \sigma_2)$ where 
\nl $\sigma_2 \notin  (A.0^+)^\om$, or $\sigma_1 \notin (A.0^+.\Si)^\om$.  

\ite $\mathcal{C}_3$ is formed by pairs  $(\sigma_1, \sigma_2)$ where 
\nl $\sigma_1 = A.w_1.A.w_2.A.w_3.A \ldots A.w_n.A.u.A.z_1 $
\nl $\sigma_2 = A.w'_1.A.w'_2.A.w'_3.A \ldots A.w'_n.A.v.A.z_2 $

\hs where $n$ is an integer $\geq 1$,  for all $i \leq n$~  $w_i, w'_i \in \Sis$,  
$z_1, z_2 \in (\Si\cup\{A\})^\om$ and 
$$u, v \in \Sis \mbox{  and }  |u| \neq |v| + 1$$

\ite $\mathcal{C}_4$ is formed by pairs  $(\sigma_1, \sigma_2)$ where 
\nl $\sigma_1 = A.w_1.A.w_2.A.w_3.A.w_4 \ldots A.w_n.A.w_{n+1}.A.v.A.z_1 $
\nl $\sigma_2 = A.w'_1.A.w'_2.A.w'_3.A.w'_4 \ldots A.w'_n.A.u.A.z_2 $

\hs where $n$ is an integer $\geq 1$,  for all $i \leq n$~  $w_i, w'_i \in \Sis$, 
$w_{n+1} \in \Sis$, 
$z_1, z_2 \in (\Si\cup\{A\})^\om$ and 
$$u, v \in \Sis \mbox{  and }  |v|\neq |u|+2$$

\end{itemize}

\noi Each set $\mathcal{C}_j$, $1\leq j\leq 4$, is easily seen to be an infinitary 
rational relation $\subseteq (\Si\cup\{A\})^\om \times (\Si\cup\{A\})^\om$ (the detailed 
proof is left to the reader).  The class $RAT_\om$ is closed under finite union thus

$$R_2 = (\Si\cup\{A\})^\om\times (\Si \cup\{A\})^\om - ( h(\Si^{\om}) \times \{\alpha\} )
 = \bigcup_{1\leq j\leq 4} \mathcal{C}_j$$

\noi is an infinitary rational relation. \ep  

\hs We can now state the following  result : 

\begin{theorem}\label{Pi} 
\noi
 For every non null countable ordinal $\gamma < \om_1^{\mathrm{CK}}$, 
there exists some  
${\bf \Si}^0_\gamma$-complete and some ${\bf \Pi}^0_\gamma$-complete
infinitary rational relations in the class $RAT_\om$.  
\end{theorem}

\proo 
 For $\gamma=1$ (and even $\gamma=2$) the result is already true 
for regular $\om$-languages. 
\nl  Let   then $\gamma \geq 2$ be a countable non null recursive 
ordinal and $L = L(\mathcal{A})\subseteq \Sio$ be a  ${\bf \Pi}_\gamma^0$-complete 
(respectively, ${\bf \Si}_\gamma^0$-complete) $\om$-language accepted by a (real time)  B\"uchi $1$-counter automaton $\mathcal{A}$. 

\hs Let $\Ga= \Si \cup \{A\}$ and $R_1 \subseteq \Gao \times \Gao$ be the infinitary rational relation constructed from $L(\mathcal{A})$ 
as in the proof of Lemma \ref{R1} and let 

$$R = R_1 \cup R_2 \subseteq \Gao \times \Gao$$
\noi The class $RAT_\om$ is closed under finite union therefore $R$ is an 
infinitary rational relation. 

\hs Lemma \ref{R1} and the definition of $R_2$ imply  that 
$R_\alpha = \{\sigma \in \Ga^\om \mid (\sigma, \alpha) \in R \}$ is equal to the set 
$\mathcal{L}= h(L)  \cup (h(\Si^{\om}))^- $ 
which   is a       ${\bf \Pi}_\gamma^0$-complete 
(respectively, ${\bf \Si}_\gamma^0$-complete)       subset of $(\Si \cup\{A\})^\om$ 
by Lemma \ref{lem}. 

\hs Moreover, for all $u\in \Ga^\om-\{\alpha\}$, 
$R_u = \{\sigma \in \Ga^\om \mid (\sigma, u) \in R \} = \Ga^\om$ holds by 
definition of $R_2$.       

\hs In order to prove that $R$ is a ${\bf \Pi}_\gamma^0$ 
(respectively, ${\bf \Si}_\gamma^0$)-complete
set remark first that 
 $R$ may be written as the  union: 
 $$R = \mathcal{L} \times \{\alpha\} ~~ \bigcup  ~~\Gao \times (\Gao - \{\alpha\})$$
\noi We already know that  $\mathcal{L}$ is a  ${\bf \Pi}_\gamma^0$ 
(respectively, ${\bf \Si}_\gamma^0$)-complete
subset of 
$(\Si\cup\{A\})^\om$. Then it is easy to show that $\mathcal{L} \times \{\alpha\}$ 
is also a  ${\bf \Pi}_\gamma^0$ 
(respectively, ${\bf \Si}_\gamma^0$)-subset 
of $(\Si\cup\{A\})^\om \times (\Si\cup\{A\})^\om$. 
 On the other side it is easy to see that  $\Gao \times (\Gao - \{\alpha\})$ is an open  
subset of $\Gao \times \Gao$.
Thus $R$ is a  ${\bf \Pi}_\gamma^0$ 
(respectively, ${\bf \Si}_\gamma^0$)-set
because the Borel class  ${\bf \Pi}_\gamma^0$ 
(respectively, ${\bf \Si}_\gamma^0$)
is closed under finite union. 
 
\hs Moreover let  $g: \Si^{\om} \ra 
 (\Si\cup\{A\})^\om \times (\Si\cup\{A\})^\om $ be the function defined  by:

$$\fa x \in \Si^{\om}~~~~~~~ g(x) = (h(x) , \alpha)$$

\noi  It is easy to see that $g$ is continuous because $h$ is continuous. By construction 
it turns out that 
for all $\om$-words $x\in \Si^{\om}$, ~~~$(x\in L)$ iff $( (h(x) , \alpha)\in R )$ iff  $(g(x)\in R)$. 
This means that $g^{-1}(R)= L$. This implies that $R$ is ${\bf \Pi}_\gamma^0$ 
(respectively, ${\bf \Si}_\gamma^0$)-complete
because $L$ is ${\bf \Pi}_\gamma^0$ 
(respectively, ${\bf \Si}_\gamma^0$)-complete.  \ep

\begin{Rem} The structure of the  infinitary rational relation $R$  can be described 
very simply by the sections $R_u$, $u\in \Gao$.  All sections but one are equal to 
$\Gao$, so they have the lowest topological complexity and exactly one section ( $R_\alpha$ ) is a 
${\bf \Pi}_\gamma^0$ 
(respectively, ${\bf \Si}_\gamma^0$)-complete
subset of $\Gao$. 
\end{Rem}

\section{Concluding remarks}

\noi The Wadge hierarchy is a great refinement of the Borel hierarchy and we have proved in \cite{cie,mscs} that the 
Wadge hierarchy of the class {\bf r}-${\bf BCL}(1)_\om$ is equal to the Wadge hierarchy of the class 
of $\om$-languages accepted by B\"uchi Turing machines. Using the above coding and similar reasoning as in  \cite{mscs}, we can easily 
infer that the Wadge hierarchy of the class $RAT_\om$ and the Wadge hierarchy of the class {\bf r}-${\bf BCL}(1)_\om$ are equal. 
Thus the  Wadge hierarchy of the class $RAT_\om$ is also the Wadge hierarchy of the (lightface)  class $\Si_1^1$ of 
$\om$-languages accepted by  Turing machines with a B\"uchi acceptance condition. In particular their Borel hierarchies are also equal. 
\nl 
We have to indicate here a mistake in  \cite{cie}. We wrote in that paper  that it is well known that if  $L \subseteq \Si^{\om}$   is a $\Si_1^1$ set 
(i.e. accepted by a Turing machine 
with a B\"uchi acceptance condition),  and is a Borel set  of  rank $\alpha$, then $\alpha$   is smaller than $\om_1^{\mathrm{CK}}$. 
This fact,  which is true if we replace $\Si_1^1$ by $\Delta_1^1$, seemed to us an obvious fact, and was  accepted by many people as true, 
but it is actually not true. 
   Kechris, Marker and Sami proved in \cite{kms} that the supremum 
of the set of Borel ranks of  (lightface) $\Pi_1^1$, so also of  (lightface) $\Si_1^1$,  sets is the ordinal $\gamma_2^1$. 
This ordinal is defined in \cite{kms} and it is proved to be strictly greater than the ordinal 
$\delta_2^1$ which is the first non $\Delta_2^1$ ordinal. Thus it holds that $ \om_1^{\mathrm{CK}} < \gamma_2^1$.  
\nl The ordinal $\gamma_2^1$ is also the supremum of the set of Borel ranks of $\om$-languages in the class {\bf r}-${\bf BCL}(1)_\om$ or in the 
class $RAT_\om$. Notice however that it is not proved in \cite{kms} that every non null ordinal $\gamma < \gamma_2^1$ is the Borel rank 
of a (lightface) $\Pi_1^1$ (or $\Si_1^1$) set, while it is known that every ordinal $\gamma < \om_1^{\mathrm{CK}}$ is the Borel rank 
of a (lightface) $\Delta_1^1$ set. 
The situation is then much more complicated than it could be expected. More details will be given in the full versions 
of \cite{cie} and of this paper.

\hs {\bf Acknowledgements.}
Thanks  to the anonymous referees for useful comments
on a preliminary version of this paper.

\end{document}